\documentclass[aps, prl, showpacs, twocolumn, amsfonts, amsmath, amssymb, superscriptaddress, floatfix]{revtex4}

\usepackage[T1]{fontenc}
\usepackage[latin9]{inputenc}
\usepackage{color}
\usepackage{graphicx}

\def\be{\begin{equation}}
\def\ee{\end{equation}}
\def\bea{\begin{eqnarray}}
\def\eea{\end{eqnarray}}

\def\me{E_c}
\def\esigma{E_{\sigma}}

\begin{document}

\title{Mobility edge for cold atoms in laser speckle potentials}

\author{Dominique Delande}
\affiliation{Laboratoire Kastler Brossel, UPMC-Paris6, ENS, CNRS; 4 Place Jussieu, F-75005 Paris, France}

\author{Giuliano Orso} 
\affiliation{Laboratoire Mat\'eriaux et Ph\'enom\`enes Quantiques, Universit\'e Paris Diderot-Paris 7 and CNRS, UMR 7162, 75205 Paris Cedex 13, France}


\date{Published in Physical Review Letters \textbf{113}, 060601 (2014)}

\begin{abstract}
Using the transfer matrix method, we numerically compute the precise position of the mobility edge of atoms exposed to a laser speckle potential, and study its dependence
vs. the disorder strength and correlation function.
Our results deviate significantly from previous theoretical estimates using an approximate self-consistent approach of localization. 
In particular we find that the position of the mobility edge in blue-detuned speckles is much lower than in
the red-detuned counterpart, pointing out the crucial role played by the 
asymmetric on-site distribution of speckle patterns. 
 
\end{abstract}

\maketitle

Anderson localization (AL), namely the absence of diffusion of waves in 
certain disordered media, is a ubiquitous phenomenon originating from the interference of multiple scatterings from random defects~\cite{Anderson:LocAnderson:PR58,Ramakrishnan:DisorderElectrons:RMP85}.
Among others, AL has been reported for
light waves in diffusive media~\cite{Wiersma:LightLoc:N97,Maret:AndersonTransLight:PRL06}
or photonic crystals~\cite{Segev:LocAnderson2DLight:N07,Lahini:AndersonLocNonlinPhotonicLattices:PRL08},
sound waves~\cite{vanTiggelen:AndersonSound:NP08},
microwaves~\cite{Chabanov:StatisticalSignaturesPhotonLoc:N00}.
Cold atomic gases do have very appealing properties for studying AL:  interference effects can be preserved over a relatively long
time, atom-atom interaction effects are small thanks to the absence
of Coulomb interaction and can be controlled using e.g. Feshbach resonances. Last, but not least, the spatial and temporal
typical scales are convenient for a direct observation of the localization effect, starting from an initially
localized wavepacket and monitoring its expansion vs. time. 

The dimension of the system is a crucial parameter. In dimension 1 (1D) and 2, AL is the generic scenario and was directly
observed on 1D atomic Bose gases~\cite{Billy:AndersonBEC1D:N08,Roati:AubryAndreBEC1D:N08}.
In dimension three (3D) and above, there is a metal-insulator transition at a specific energy $E\!=\!\me$, the so-called mobility edge, 
separating localized $(E<\me)$ from diffusive $(E>\me)$ states. 

The kicked rotor model describing cold atoms
exposed to a periodic or quasi-periodic sequence of far-detuned laser pulses can be mapped to an 
Anderson model in momentum space, making it possible to observe 1D AL in  momentum space~\cite{Moore:AtomOpticsRealizationQKR:PRL95} in 1995. More recently, the 3D Anderson transition
has been experimentally observed~\cite{Chabe:Anderson:PRL08}, the critical exponent accurately measured and its universality tested~\cite{Lopez:ExperimentalTestOfUniversality:PRL12}.

Experiments ~\cite{Kondov:ThreeDimensionalAnderson:S11,Jendrzejewski:AndersonLoc3D:NP12,Semeghini:2014} on 3D AL in configuration space, using laser speckles as a source of disorder for atoms, are less advanced. 
The reasons are twofold. First the broad-band energy distribution of the expanding
atoms severely complicates the extraction of the mobility edge from the raw data. 
Second and most important, a direct comparison with theory remains problematic. 
To date,  there is indeed no \emph{exact} theoretical prediction for the mobility edge of atoms in speckle potentials.
The currently available~\cite{Kuhn:Speckle:NJP07,Yedjour:MobilityEdge3D:EPJD10,Piraud:MobilityEdge3D:NJP13}  estimates 
are based on an approximate self-consistent theory of localization (SCTL)~\cite{Vollhardt:SelfConsistentTheoryAnderson:92}.

The experimentally measured~\cite{Kondov:ThreeDimensionalAnderson:S11} position of the mobility edge  in highly anisotropic speckles is unexpectedly high compared to the SCTL results. 
In Ref.~\cite{Jendrzejewski:AndersonLoc3D:NP12}, the experimentally measured mobility edge is significantly below the most naive implementation of the SCTL~\cite{Kuhn:Speckle:NJP07}. A modified version of the SCTL~\cite{Piraud:EPL:2012} improves that agreement, with however a predicted mobility edge below the measured value. 
In Ref.~\cite{Semeghini:2014}, the authors developed a new experimental scheme to directly measure the mobility edge of atoms.  For relatively small disorder, their result is apparently above the average potential value, in contrast with the SCTL theoretical 
expectations. 

Contrary to typical condensed matter systems, the on-site distribution $P(V)$ of the blue-detuned speckles 
employed in experiments~\cite{Kondov:ThreeDimensionalAnderson:S11,Jendrzejewski:AndersonLoc3D:NP12,Semeghini:2014} is not Gaussian,
but follows a Poissonian law~\cite{Kuhn:PRLSpeckle:2005,Goodman:07}:
\begin{equation}
 P(V) = \frac{\Theta(V+V_0)}{V_0}\ \exp{\left(-\frac{V+V_0}{V_0}\right)}
\label{Eq:Rayleigh-up}
\end{equation}
where  $\Theta$ is the Heaviside function and $V_0$ is related to the variance by  $\langle V^2\rangle\!=\!V_0^2$. 
For convenience, we have shifted the potential by its average value. 
Red-detuned speckles are described by Eq.(\ref{Eq:Rayleigh-up}) under the change $V\rightarrow -V$.

In this Letter we use transfer matrix calculations together with the finite-size scaling to pinpoint the precise position of the mobility edge for massive particles in laser speckle potentials. We show that {\sl i)}  for blue-detuned speckle, the previous theoretical calculations
based on the SCTL significantly overestimate the correct position of the mobility edge;  {\sl ii)} the values of $E_c$ 
for red and blue speckles are completely different due to the asymmetry $P(V)\neq P(-V)$ of the  distribution  
(\ref{Eq:Rayleigh-up}), while the current implementations of the SCTL would predict the same value;
{\sl iii)} at least for  isotropic speckles,  the position of the mobility edge depends mainly on the width
of the correlation function $\langle V(0)V(\mathbf r)\rangle$ of the potential rather than on its specific shape. 

\begin{figure}
\includegraphics[width=0.9\columnwidth]{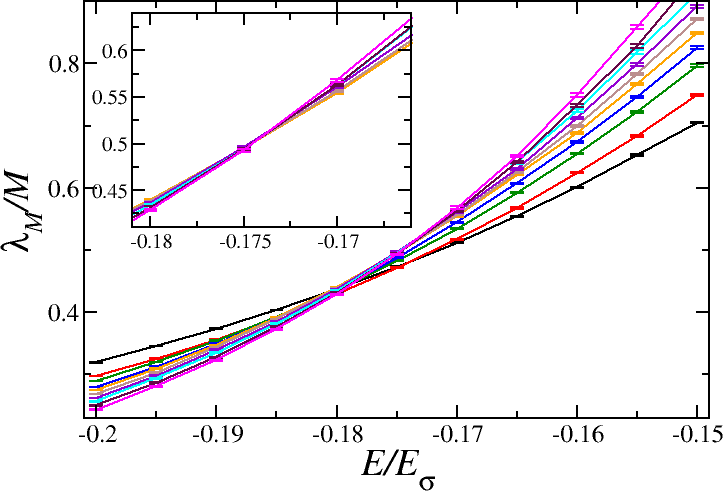}
\caption{(color online) Numerical determination of the mobility edge of a cold atom in a disordered 3D potential created
by a blue-detuned speckle with scaled strength $V_0\!=\!0.5\esigma.$ For a given energy, we compute the localization length
$\lambda_M$ of a long bar-shaped grid with square section $M\!\times\!M$. 
Each curve is computed for a single $M$ value
and shows $\lambda_M/M$ vs. energy. The various curves cross at the position of the mobility edge. The main plot includes
curves from $M\!=\!35$ (less steep, black curve) to $M\!=\!80$ (steepest magenta curve), that suggest a mobility edge around $-0.18\esigma$. When only the largest $M$ values are kept -- in the inset -- a better estimate $E_c\!=\!-0.173\esigma$ is obtained.}
\label{Fig:Crossing}
\end{figure} 

The transfer-matrix approach was first used to study the 3D 
 Anderson model~\cite{McKinnonKramer:TransferMatrix:ZPB83,Slevin:CriticalExponent:NJP2014}. 
Its application to speckle patterns is challenging because 
the spatial correlations of the potential requires very dense grids, increasing drastically the computational effort.
Our starting point is to spatially discretize the Schr\"odinger equation on a cubic grid with step $\Delta$: the 
Laplace operator is replaced by a 7 points sum corresponding to an effective hopping rate 
$J\!\!=\!\!\hbar^2/(2m\Delta^2).$ 
The disordered potential at each site is generated by properly convoluting a $\delta$-correlated random sequence
in order to reproduce both the on-site potential distribution $P(V)$ and the desired correlation function~\cite{Goodman:07,Kuhn:Thesis:07}, see Supplemental Material.
We first consider the case of a blue speckle, whose distribution is given by Eq.~(\ref{Eq:Rayleigh-up}). Following ~\cite{Kuhn:Speckle:NJP07,Yedjour:MobilityEdge3D:EPJD10,Piraud:MobilityEdge3D:NJP13}, 
we chose the simplest case of a speckle created by a statistically isotropic 3D illumination, with correlation function:
\begin{equation}
 \langle V(\mathbf{r'})V(\mathbf{r'+r}) \rangle = V_0^2 C(r/\sigma)
\label{Eq:Correlation_Speckle}
\end{equation}
where $C(r)\!=\!(\sin r/r)^2$  and $\sigma$ is the characteristic correlation length. Notice that this is different from the
scale-free disorder case investigated in~\cite{Ndawana:ScaleFreeDis:2004}. The associated ''correlation'' energy  sets the important energy scale:
\begin{equation}
\esigma = \frac{\hbar^2}{m\sigma^2}
\label{Eq:esigma}
\end{equation}
where $m$ is the mass of the atom. In the following we measure all energies in units of $\esigma.$ Varying $V_0/\esigma$ can be equivalenty viewed as varying the disorder strength $V_0$ at fixed $\sigma$ or the correlation length $\sigma$
at fixed $V_0.$
 
The discretization step must be chosen sufficiently small to resolve
the details of the disordered potential and the oscillations of the wavefunction, that is smaller than $\sigma$ and the de Broglie wavelength, which are similar when $V_0$ and $E$ are of the order of unity. We have thus chosen a 
discretization step $\Delta\!=\!0.2\pi\sigma.$ We have carefully checked that, up to $V_0\!=\!1,$ the discretization error on the determination of the mobility edge is smaller than $0.005.$

The transfer matrix method is used to recursively compute the total transmission
of a bar-shaped grid with length $L$ and square transverse section $M\!\times\!M$, with $M\!\ll\!L$.
 This system can be viewed as quasi-1D and is thus 
Anderson localized: its total transmission 
decays like $\exp(-2L/\lambda_M)$ where $\lambda_M$ is the quasi-1D localization length in units of the lattice spacing. In practice, the log of the total transmission is a self-averaging quantity
which can be safely computed, together with error bars, by using either very long bars 
or smaller ones but averaging over many independent realizations of the disorder.
Periodic boundary conditions are used in the transverse directions to reduce finite-size effects. 
 
$\lambda_M$ depends on $M,$ on the energy and on the disorder strength. 
Figure~\ref{Fig:Crossing} shows the ratio $\lambda_M/M$ as a function
of energy, for various $M$ values, at a fixed disorder strength $V_0\!=\!0.5.$
At low energy, in the localized regime, $\lambda_M/M$ decreases with $M$ and eventually behaves like 
$\lambda_{\infty}/M$, with $\lambda_{\infty}\!=\!\lim_{M\to\infty} \lambda_M$ the 
3D localization length in units of $\Delta$.
In contrast, at high energy,  $\lambda_M/M$ increases with $M$, a signature of the diffusive regime.
At the mobility edge, $\lambda_M/M$ is a constant $\Lambda_c$ of order unity, meaning that the quasi-1D localization length is comparable to the transverse size of the system, a signature of marginal 3D localization. 
Thus, the mobility edge can be obtained by looking at the point where all curves cross in Fig.~\ref{Fig:Crossing}.  For small $M,$ finite-size effects make the crossing not perfect: looking at only the lowest $M$, one could get the false impression that the crossing takes place
at $E\!=\!-0.18,$ while the highest $M$ values (inset) show that the true crossing is at  $E\!=\!-0.173\pm0.002.$ 
Reaching such a high precision in the determination of the mobility edge requires massive computing
resources: transverse system size up to $M\!=\!80$, longitudinal size of the bar up to 1 million sites. Altogether, data shown in Fig.~\ref{Fig:Crossing} required 200 000 hours of computation on a supercomputer
with slightly more than $10^{19}$ arithmetic operations.

Finite-size scaling is a technique making it possible to go beyond the visual detection of the crossing, 
pinpointing more accurately the position of the mobility edge as well as the critical exponent of the Anderson transition.
The drift of the crossing point in Fig.~\ref{Fig:Crossing} is due to the presence of damped oscillations in
 $\lambda_M$ as $M$ increases. These oscillations, which are absent  in the
 Anderson model with uncorrelated disorder~\cite{Slevin:CriticalExponent:NJP2014}, are probably
 related to the tail of the correlation function $C(r)$ for $r\approx M\Delta$ and can be hardly 
incorporated in standard finite-size scaling techniques~\cite{Slevin:CriticalExponent:NJP2014}  as ``irrelevant variables''. 
 We have used this technique, retaining only the data with $M\geq 50$,  to confirm the position of the mobility edge at $E_c\!=\!-0.172\pm0.0015,$ with a critical $\Lambda_c\!=\!\lim_{M\to\infty}\lambda_M/M\!=\!0.54\pm 0.03$ 
and $\nu\!=\!1.6\pm0.2$. These results are less accurate but compatible with the values
$\Lambda_c\!=\!0.5765\pm 0.001$ and $\nu\!=\!1.571\pm0.008$ obtained for the Anderson model~\cite{Slevin:CriticalExponent:NJP2014}. 
Details of the calculation are given in the Supplemental Material, together with a thorough analysis of the
single parameter scaling.

\begin{figure}
\includegraphics[width=0.9\columnwidth]{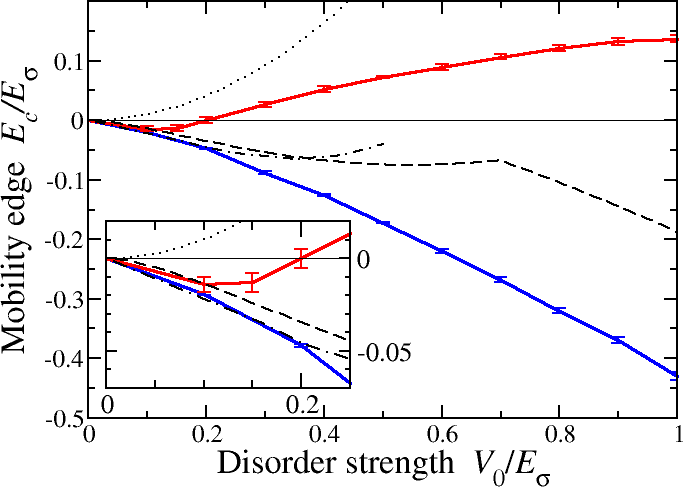}
\caption{(color online) Mobility edge separating localized atoms at low energy from diffusive atoms at high energy for cold atoms in a disordered 3D speckle potential 
(energies in units of the correlation energy, Eq.~(\ref{Eq:esigma})).
The solid lines (lower, blue curve and upper, red curve) correspond respectively to blue and red detuned speckles. While the
former mobility edge lies always below the average potential (shown by the horizontal thin line), the latter becomes positive as the potential strength increases (see zoom of the small $V_0$ region in the inset). 
The dotted line shows the prediction of the naive self-consistent theory of localization~\cite{Kuhn:Speckle:NJP07}. The dashed~\cite{Piraud:MobilityEdge3D:NJP13} and dash-dotted~\cite{Yedjour:MobilityEdge3D:EPJD10} lines are the predictions 
of improved self-consistent theories incorporating the real part of the self-energy
which nevertheless lie rather far from the exact numerical results and do not describe the large difference between blue and red speckles.}
\label{Fig:Mobility_Edge}
\end{figure}

We repeat the same calculation for various values of the potential strength. To save computer resources,
we use smaller system sizes, producing larger finite size effects, thus slightly larger error bars on the position of the mobility edge, but always smaller than $0.01\esigma.$  
The obtained results for the blue speckle are shown in Fig.~\ref{Fig:Mobility_Edge}  by the blue solid line. The mobility edge is negative for all  disorder strength $V_0$;
in other words, the mobility edge always lies \emph{below} the average potential.

We also perform analogous calculations for a red speckle, but keeping the spatial correlation function of the 
 disorder, Eq.~(\ref{Eq:Correlation_Speckle}), unchanged. This is shown in Fig.~\ref{Fig:Mobility_Edge}  by the red solid line. 
For small disorder strengths,  the mobility edges of red and blue speckles are both negative and very close to each other. In this deep ``quantum'' regime, the
de Broglie wavelength around the mobility edge is larger than $\sigma$: the quantum particle thus averages the local potential fluctuations,
and is less sensitive to the skewness of the potential distribution.  For stronger disorder $V_0\gtrsim 0.2,$ the effect of the asymmetric potential is more important,
the mobility edge of the red speckle becomes positive (as clearly shown in the figure inset, despite rather large error bars) and lies well above the blue one. Significant blue/red differences
in localization properties have already been observed in the 1D case~\cite{Gurevich:Lyapounov1D:PRA2009,Lugan:Lyapounov1D:PRA2009}.

We now compare our precise numerical results with the three available theoretical estimates based on 
different implementations of the SCTL. These are shown in Fig.~\ref{Fig:Mobility_Edge} by the dotted~\cite{Kuhn:Speckle:NJP07},  
the dash-dotted~\cite{Yedjour:MobilityEdge3D:EPJD10} and the dashed~\cite{Piraud:MobilityEdge3D:NJP13} lines.
Despite the SCTL being an approximate theory, it turned out to be quite good for simple models such as the Anderson model~\cite{Kroha:SelfConsistentTheoryAnderson:PRB90}.
The SCTL was naively applied to speckles in~\cite{Kuhn:Speckle:NJP07} assuming that the spectral function and the density of states are unaffected by the disorder. The prediction of this (on-shell) approximation
is that the mobility edge is always positive, see Fig.~\ref{Fig:Mobility_Edge}, and is obviously rather bad. 

A simple attempt to cure this problem is to incorporate in the SCTL the real part of the self-energy, which shifts the effective lower bound of the spectrum to negative energies~\footnote{For a blue speckle, there is a strict lower bound $-V_0$ for the energy spectrum, but the density of states is vanishingly small just above $-V_0.$  For a red speckle, the energy spectrum is unbounded.}. 
Two 
variants of this simple idea have been used~\cite{Yedjour:MobilityEdge3D:EPJD10,Piraud:MobilityEdge3D:NJP13}, predicting that the 
mobility edge of the blue speckle is actually negative.  However  we see in Fig.~\ref{Fig:Mobility_Edge} that these theories become
rapidly inaccurate as the disorder strength increases. First, they significantly overestimate the mobility edge. Second, in these approaches, the scattering amplitude is still evaluated at the lowest order in the Born approximation, making the mobility edge dependent only on the correlation function, Eq.~(\ref{Eq:Correlation_Speckle}), but not on the  potential distribution,  Eq.~(\ref{Eq:Rayleigh-up}). As a consequence the SCTL predicts 
\emph{identical} mobility edges for blue and red speckles, which is clearly incorrect. This is not surprising, as the SCTL is not aimed at producing quantitative results
beyond the simplest (Gaussian) disorder models.

\begin{table}[t]
\begin{centering}
\begin{tabular}{|p{3.4cm}|p{2.0cm}|p{2.0cm}|}
\hline
 Correlation function $C(r)$, eq.~(\ref{Eq:Correlation_Speckle})& Blue-detuned mobility edge  & Red-detuned mobility edge  \tabularnewline
\hline
\hline
$\left[\sin r/r\right]^2$ & -0.172(2) & 0.073(2)  \tabularnewline
\hline
$\exp(-r^2/2)$ & -0.181(4) & 0.054(4) \tabularnewline
\hline
$\left[3(\sin r/r -\cos r)/r^2\right]^2$ & -0.175(5) & 0.053(8) \tabularnewline
\hline

\end{tabular}
\end{centering}

\caption{\label{Table:Mobility_Edge} Mobility edge for various disordered speckle potentials, all calculated for
$V_0\!=\!0.5\esigma .$ The various potentials correspond to different spatial correlation functions, adjusted
to have the same width at half-maximum. The mobility edge
is computed both for the blue-detuned (positive detuning) and the red-detuned (negative detuning) cases. The digit in parenthesis
is the uncertainty on the last digit.
The most important factor is the sign of the detuning, while the precise form
of the spatial correlation function has only a small effect on the position of the mobility edge.}
\end{table}

\begin{table}[t]
\begin{centering}
\begin{tabular}{|c|c|}
\hline
 Correlation function $C(r)$& Mobility edge\tabularnewline
\hline
\hline
$\left[\sin r/r\right]^2$ & -0.135(5) \tabularnewline
\hline
$\exp(-r^2/2)$ & -0.124(4) \tabularnewline
\hline

\end{tabular}
\end{centering}

\caption{\label{Table:Mobility_Edge2} Mobility edge for Gaussian distributed disordered potentials, calculated for
$V_0\!=\!0.5\esigma,$ and two different spatial correlation functions. The mobility edge is between the blue and red results
in table~\ref{Table:Mobility_Edge}, confirming that the potential distribution is the important factor. Conversely, the precise form
of the spatial correlation function has only a small effect.}
\end{table}

\begin{figure}
\includegraphics[width=0.9\columnwidth]{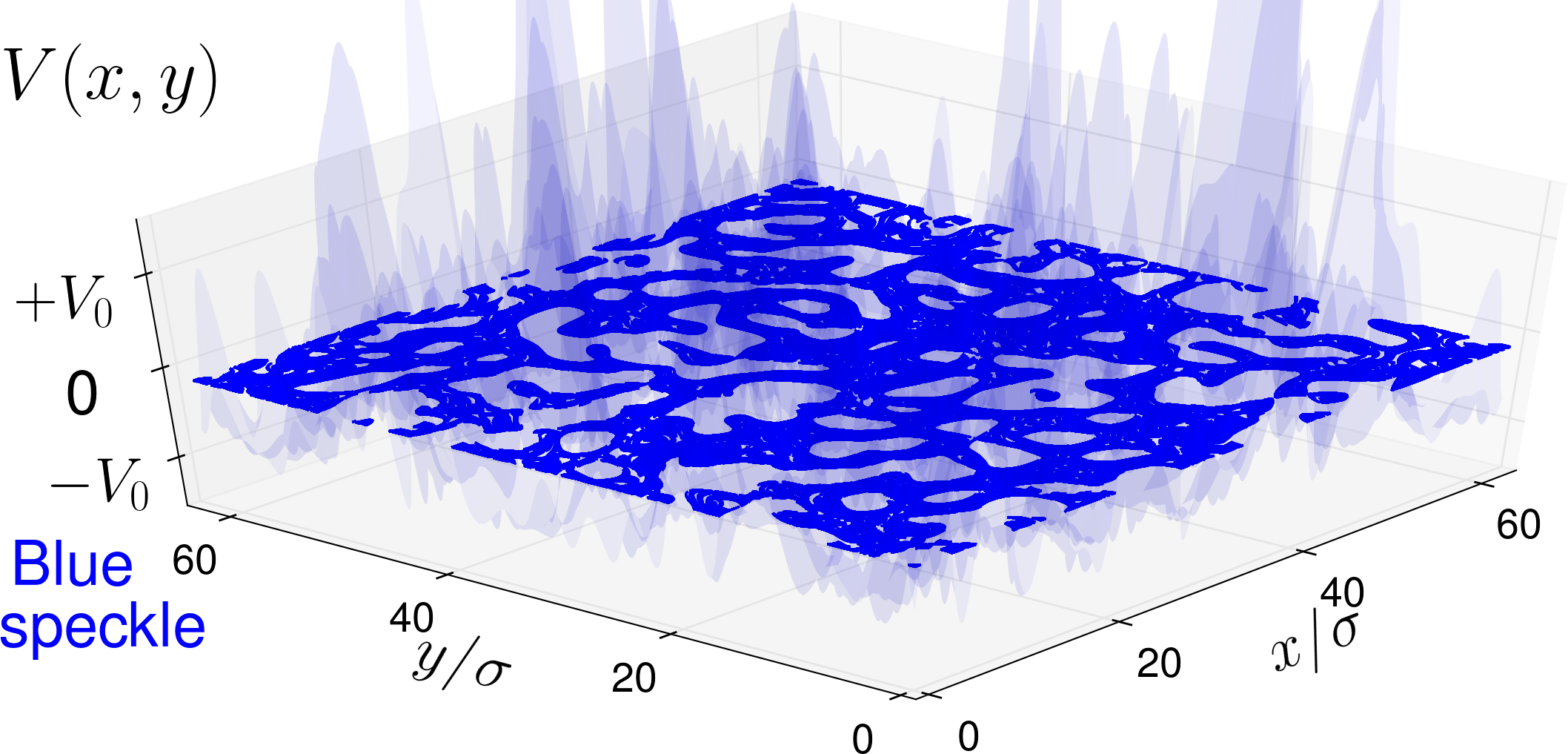}
\includegraphics[width=0.9\columnwidth]{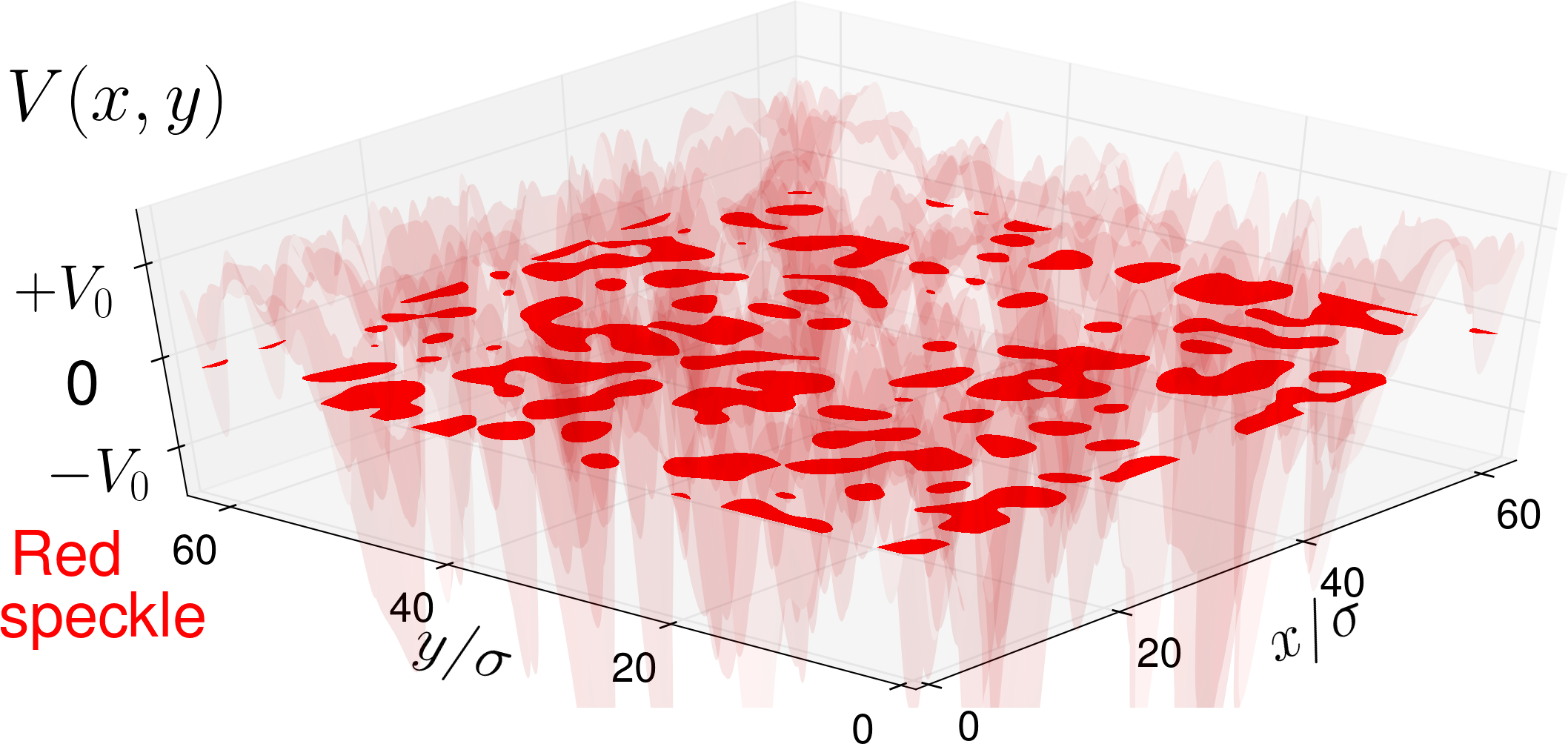}
\caption{(color online) Planar cuts in color: classically allowed regions in a transverse section of the grid 
at energy $-0.15 V_0$ (transparent regions are classically forbidden). The allowed region is well connected for the blue speckle and disconnected for the red speckle.}
\label{Fig:Blue_red_difference}
\end{figure} 

For further investigation, we use different models of disordered potentials. 
Table~\ref{Table:Mobility_Edge} shows the mobility edges at $V_0=0.5,$ for red and blue speckles, and various 
isotropic correlation functions $C(r)$ (the third one was adopted in Refs.\cite{Pilati:MonteCarlo:PRL2009,Pilati:MonteCarlo:NJP2010}). 
In order to make the comparison meaningful, we have adjusted the
parameter $\sigma$ for the potential correlation function to reproduce the same width at half-maximum in the 3 cases, keeping
the same strength $V_0$ of the potential.
The mobility edges are almost identical for the 3 blue speckles, around -0.175,
and almost identical for the 3 red speckles, around +0.06. This proves that the essential ingredient which determines
the mobility edge is the distribution $P(V)$ of disorder strength, and more specifically the fact that the distribution, eq.~(\ref{Eq:Rayleigh-up}), is very asymmetric. For a blue-detuned speckle, there is a strict lower bound at $-V_0$ (corresponding to the nodes of the speckle) and a long tail towards high $V$ (bright spots of the speckle).
There is thus a large probability to find the potential between $-V_0$ and $0.$ For a red detuned speckle,
the potential is upside-down, so that the largest probability is to have the potential between $0$ and $+V_0$ and a long tail
towards negative $V$ (in both cases,
the average potential is 0).

Figure~\ref{Fig:Blue_red_difference} shows the classically available regions on a transverse section of the grid, $V(x,y,z\!\!=\!\!z_0)\!<\!E$,
for blue and red speckles at the same energy $E\!=\!-0.15V_0,$ roughly half-way between the two mobility edges. The blue region is strongly connected -- which favors delocalization -- 
while the red region is composed of disconnected islands. This percolation argument is mostly valid in the semiclassical limit where $V_0\gg E_{\sigma},$ but it already gives the qualitative trend at low $V_0$ where tunneling plays an important role~\cite{Shapiro:JPA:2012}.  

The details of the potential correlation function just give rise
to small variations of the mobility edge. At the mobility edge, the disorder is so strong that all orders of the Born expansion contribute to the transport properties, smoothing out all details. For example, the kink around $V_0\!=\!0.7$ 
in the dashed line of Fig.~(\ref{Fig:Mobility_Edge}), due to the specificity of the potential correlation function ~\cite{Piraud:MobilityEdge3D:NJP13}, is no longer visible in the ``exact'' numerical results.

For a symmetric Gaussian distribution of potential $P(V)\!\!=\!\!\exp(-V^2/2V_0^2)/\sqrt{2\pi V_0^2},$ the mobility edge, shown in table~\ref{Table:Mobility_Edge2}, is similar for various correlation functions, around -0.13, but significantly different
from the blue and red speckles, although the current implementation of the SCTL gives exactly the same
value for the three distributions, around -0.074.

The SCTL does not make accurate quantitative predictions because it uses the lowest order in the Born approximation. As shown in~\cite{Shapiro:JPA:2012},
it already does not correctly predict the density of states.
The large red-blue difference unambiguously demonstrates that odd powers of $V_0$ (absent for a Gaussian distribution) play a major role. Even when such terms are absent, the SCTL is not very accurate. We think that inclusion of higher
order terms -- even approximately -- is a necessary step to make these theories more quantitative.

To summarize, we have shown that it is possible to numerically compute the mobility edge for cold atoms exposed
to a disordered potential created by a 3D optical speckle. The calculation is quasi-exact (limited only by computer resources) 
and its accuracy largely sufficient for comparison with experimental results. The mobility edge for a blue-detuned speckle
is significantly lower than predicted by the self-consistent theory of localization. We attribute the difference
to the peculiar potential distribution of the speckle, while the spatial correlation function defines the characteristic
energy scale $\esigma$ but seems otherwise to play a rather minor role.
 While we have used  for simplicity an isotropic disorder correlation function, the method can be  extended
to the more anisotropic configurations used in the experiments. In the view of the
present results, it seems likely that the actual mobility edge lies somewhat lower
than measured in the experiments using blue-detuned speckles~\cite{Kondov:ThreeDimensionalAnderson:S11,Jendrzejewski:AndersonLoc3D:NP12,Semeghini:2014}. This problem is currently under investigation. 

We thank N. Cherroret, C. M\"uller, B. Shapiro and S. Pilati for useful discussions.  This work was granted access to the HPC resources of TGCC under the allocation 2013-056089 made by GENCI (Grand Equipement National de Calcul Intensif) and to the HPC resources of The Institute for scientific Computing and Simulation financed by Region Ile de France and the project Equip@Meso (reference ANR-10-EQPX- 29-01).

\bibliographystyle{apsrev}
\bibliography{ArtDataBase}

\end{document}


\title{Supplemental material for Mobility edge for cold atoms in laser speckle potential}

\maketitle

 \vspace{0.5cm}
 \textbf{I. Numerical generation of speckle potentials}  
 \vspace{0.5cm}
 
In a real experiment, the optical speckle potential is created by superimposing plane waves coming from different
 directions with random phases.  The electric field can then be written as:
\begin{equation}
 E(\vecr) = \sum_{\veck_i}{A_i \rme^{\rmi \veck_i\cdot \vecr}}W(\veck_i),
\label{eq:Esum}
\end{equation}
where $A_i$ is a random complex number with unit modulus, representing the random phase of the plane wave with wave vector $\veck_i,$
and $W(\mathbf k_i)$ is the corresponding weight.  
 
If a large number of uncorrelated plane waves is used, as a consequence of the central limit theorem, 
the real and the imaginary parts of the electric field at a given point are uncorrelated and follow 
a Gaussian distribution~\cite{Kuhn:Speckle:NJP07,Shapiro:JPA:2012}.
The optical potential felt by the atom is the modulus square of the electric field, so the 
on-site potential distribution $P(V)$ follows the Rayleigh law (Eq.~(1) of the Letter).

As shown in~\cite{Kuhn:Speckle:NJP07}, the spatial correlation function of the optical potential is nothing but the squared modulus
of the correlation function of the electric field. The latter is directly related to the distribution of the $\veck_i$ vectors in the sum, eq.~(\ref{eq:Esum}). If the $\veck_i$ are chosen isotropically on the surface of a sphere, 
$W(k)=\delta(|\vec k|-k_0) $, the field correlation function
is a simple sinc function, corresponding to $C(r)=[\sin r/r]^2$ in Eq.~(2) of the Letter, with $\sigma=1/k_0.$ If the wave-vectors are chosen with a uniform density inside a sphere, 
$W(k)=\Theta(k_0-|\vec k|) $, a simple calculation shows that one has: $C(r)=[3(\sin r/r-\cos r)/r^2]^2.$  
Finally one produces a Gaussian spatial correlation function using $W(k)=\exp(-k^2/2k_0^2).$

Since the position of the mobility edge is calculated numerically via  the transfer-matrix method, we consider
a bar-shaped grid with a square transverse section $M\times M$ and a length $L\gg M.$
 In order to resolve the spatial variations of the potential, the 
grid step must be small compared to the correlation length of the potential, $\Delta \ll \sigma$.

In order to minimize finite size effects, we chose to use periodic boundary conditions in the two transverse directions.
This implies that the disordered potential has the same transverse periodicity. This is of course incompatible with the
decay at infinity of the correlation function $C(r),$ Eq.~(2) of the Letter. However, converging results are obtained if
the transverse size of the system if much larger than the correlation length of the potential, that is $M\Delta\gg\sigma.$
Thus, too small values of $M$ must be excluded (in practice, we use at least $M=30$, that is $M\Delta=6\pi\sigma$) which explains why the calculation requires large computer resources.

\begin{figure}
\includegraphics[width=0.9\columnwidth]{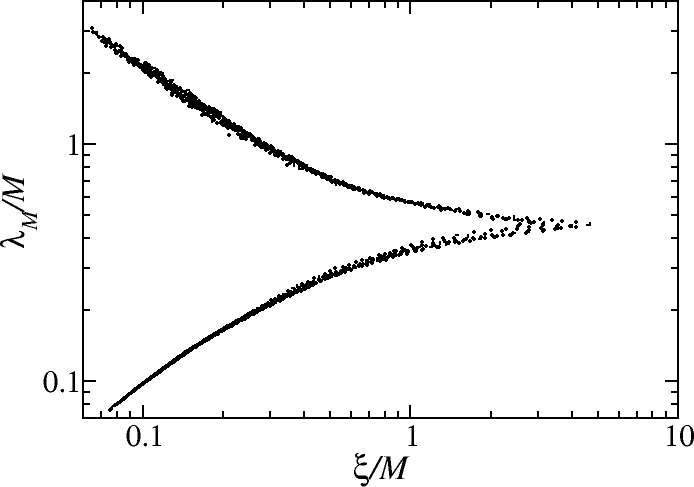}
\caption{Scaling function numerically obtained for cold atoms in a disordered 3D blue-detuned speckle potential
with strength $V_0=0.5\esigma.$ Data for 201 values of the energy in the range [-0.27,-0.07] and 6 transverse system sizes
$(M=30,35,40,45,50,55)$ collapse on a unique two-valued scaling function. The lower (resp. upper) branch corresponds to the
localized (resp. diffusive) side of the transition; the tip is associated with the divergence
of the localization length $\xi(E)$ at the critical point.}
\label{fig:scaling_function}
\end{figure}

\begin{figure}
\includegraphics[width=0.9\columnwidth]{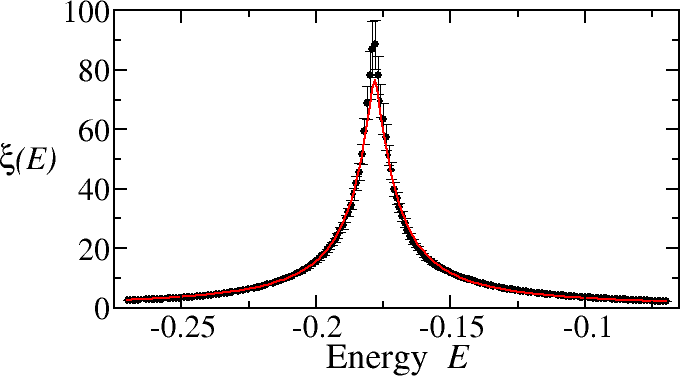}
\caption{Localization (resp. correlation) length $\xi(E)$ on the localized (resp. diffusive) side of the metal insulator Anderson
transition for cold atoms in a disordered 3D blue-detuned speckle potential
with strength $V_0=0.5\esigma$ (same data than in figure 1). The algebraic divergence near the critical energy $E_c\approx-0.178$
is a characteristic feature of the Anderson transition. The points with error bars are the numerical results
and the solid red curve the fit using an algebraic dependence with a cut-off, Eq.~(\ref{eq:xi}).
}
\label{fig:xi}
\end{figure}

\vspace{0.5cm}
 \textbf{II. Finite-size scaling}  
 \vspace{0.5cm}

\textsl{Test of one-parameter scaling}.
As a test of the one-parameter scaling properties of the observed Anderson transition, we used the original finite-size approach
described in~\cite{McKinnonKramer:TransferMatrix:ZPB83}. 
For this purpose, we gather the quasi-1D localization length $\lambda_M$ determined for various values of $M$ and of the energy $E$ (at fixed disorder strength). The one-parameter scaling hypothesis is that there is a unique scaling function $\cal{F}$ -- or rather
a function with two branches -- such that:
\begin{equation}
 \frac{\lambda_M(E)}{M} = {\cal F} \left(\frac{\xi(E)}{M}\right),
\end{equation}
where $\xi(E)$ is a fitted characteristic length proportional to the localization length in the localized regime
and to the inverse of the diffusion constant in the diffusive regime. 

Figure~\ref{fig:scaling_function} shows the scaling function obtained for $V_0=0.5.$ Since no assumption is made concerning the
specific form of the scaling function ${\cal F}(x)$, we have used hundreds energy values and, consequently, 
relatively small systems sizes, $M\leq 55$.
The lower (resp. upper) branch is the localized
(resp. diffusive) side of the Anderson transition, and the tip on the right side of the figure shows the divergence
of the localization length $\xi(E)$ near the critical point. 

Figure~\ref{fig:xi} shows $\xi(E).$ Because the system is finite, it cannot display an infinite divergence. In order to take into
account the unavoidable existence of cut-offs, we fit the results with the following formula:
\begin{equation}
 \xi(E) = \frac{\alpha}{\beta+|E-E_c|^\nu}
\label{eq:xi}
\end{equation}
where $E_c$ is the mobility edge, $\nu$ the critical exponent, $\alpha$ a simple multiplicative factor and
$\beta$ a cut-off.

The fit, shown in Fig~\ref{fig:xi}, is good, although there are clearly deviations in the tails, which is not
surprizing as Eq.~(\ref{fig:xi}) is expected to be valid only in the very vicinity of the critical point.
From the data of Fig.s~\ref{fig:scaling_function} and \ref{fig:xi}, we extract $E_c=-0.178$, $\Lambda_c=0.46$ and
$\nu=1.4$. These values are not very accurate, being affected by strong finite-size corrections,  drifting
the crossing point towards $-0.18$, as shown in Fig.1 of the Letter. 

\textsl{Scaling ansatz.}  More accurate results can be obtained focusing exclusively on large system sizes,
for which damped oscillations in $\lambda_M$ are less important. We thus apply the finite-size scaling ansatz to the numerical 
data shown in Fig.1 of the Letter with $M\geq 50$ following the procedure outlined in Ref.~\cite{Slevin:CriticalExponent:NJP14}.

\begin{figure}[tb]
\includegraphics[width=0.9\columnwidth]{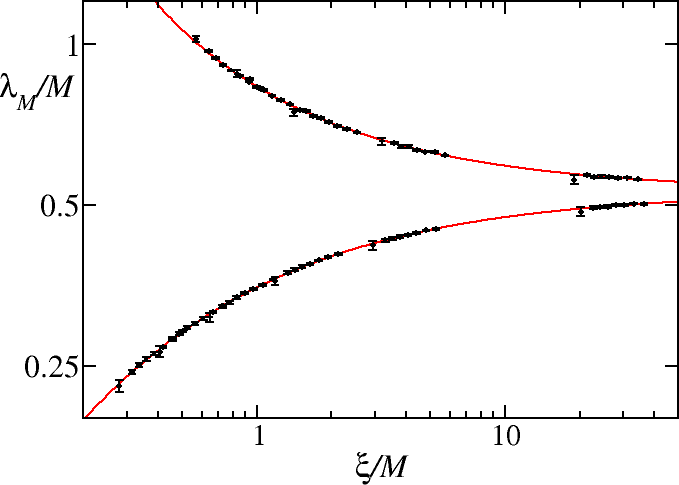}
\caption{Same scaling function of Fig.\ref{fig:scaling_function} but this time obtained using the numerical data in Fig.1 of the Letter with $50 \leq M\leq 90$ and the Taylor expansion of the ansatz. We obtain $E_c=-0.172\pm0.0015$, $\Lambda_c=0.54\pm0.03$ and $\nu=1.6\pm0.2.$, compatible with the values $\Lambda_c=0.576$ and $\nu=1.57$ found for the 3D Anderson model at zero energy~\cite{Slevin:CriticalExponent:NJP14}. The error bars are comparable to the size of the symbols, except
for $M\!=\!90$ (most left point in each series) where there are larger because of less configuration averaging. Altogether,
the quality of the fit is satisfactory with a $\chi^2$ per degree of freedom equal to 1.63.
 }
\label{fig:better}
\end{figure}

Taking into account  the possible existence of irrelevant operators, the ratio 
$\lambda_M/M$ can be written in terms of a new scaling function $\tilde {\cal F}$ as
\begin{equation}\label{slevin1}
 \frac{\lambda_M(E)}{M} = \tilde {\cal F} (\phi_1,\phi_2),
\end{equation}
where the arguments $\phi_1$ and $\phi_2$ correspond, respectively, to relevant and irrelevant scaling variables. 
By setting $\omega=(E-E_c)/E_c$, the latter can be expressed as
\begin{equation}\label{slevin2}
\phi_1=u_1(\omega) M^{1/\nu}\;\;\;\,\phi_2=u_2(\omega) M^{y},
\end{equation}
where $y<0$ is the exponent of the (dominant) irrelevant operator. In proximity of the critical point, the functions $u_i(\omega)$ 
are Taylor-expanded up to the order $m_i$:
\begin{equation}\label{slevin3}
u_i(\omega)=\sum_{j=0}^{m_i} b_{i,j} \omega^{j},
\end{equation}
where by construction $b_{1,0}=0$. Next the scaling function $\tilde {\cal F}(\phi_1,\phi_2)$ is also Taylor-expanded up
to orders $n_1$ and $n_2$, respectively, yielding
\begin{equation}\label{slevin4}
\tilde {\cal F}(\phi_1,\phi_2)=\sum_{j_1=0}^{n_1}\sum_{j_2=0}^{n_2} a_{j_1,j_2} \phi_1^{j_1}\phi_2^{j_2}.
\end{equation}
To avoid ambiguity in the fitting model we need to set $a_{1,0}=a_{0,1}=1$~\cite{Slevin:CriticalExponent:NJP14}.

We have fitted our reduced data set using Eqs (\ref{slevin1}-\ref{slevin4}). 
A key point is the choice of the parameters $(n_i,m_i).$ Following~\cite{Slevin:CriticalExponent:NJP14}, we choose these parameters just sufficiently
large to ensure the fit to be satisfactory, that is with a $\chi^2$ per degree of freedom of the order of unity.
Because of (damped) oscillatory behavior with increasing $M$ -- not taken into account in the algebraic irrelevant variables -- some
deviation from the best fit remains, but is typically limited to a $\chi^2$ per degree of freedom in the range from 1 to 4.
With $n_1\!=\!m_1\!=\!2$ and $n_2\!=\!1,m_2\!=\!0$, we obtain $E_c=-0.172\pm0.0015$, $\Lambda_c=0.54\pm0.03$ and $\nu=1.6\pm0.2.$ 
Similar results are obtained when the degrees parameters $(n_i,m_i)$ are varied. 

The obtained results are compatible with the values $\Lambda_c=0.5765\pm 0.001$ and $\nu=1.571\pm0.008$ found \cite{Slevin:CriticalExponent:NJP14} for the 3D Anderson model at zero energy. 
It is not clear to us whether exactly the same value of $\Lambda_c$ should be obtained: indeed, the Anderson model has an anisotropic dispersion relation in the absence of disorder, especially near the band center where the mobility edge is usually studied.
This could in principle affect the value of $\Lambda_c$.

The more accurate scaling function is then shown in Fig.~\ref{fig:better} where we made use of the relation $\xi(E)= |u_1(\omega)|^{-\nu}$. No siginificant deviation of the data from the fit is visible.

\bibliographystyle{apsrev}
\bibliography{ArtDataBasev4}